\documentclass[prd,onecolumn]{revtex4}
\usepackage{dcolumn}
\usepackage{multirow}
\usepackage{graphicx}
\usepackage{amssymb}
\usepackage{bm}
\usepackage{hyperref}
\usepackage{epstopdf}
\usepackage{color}
\usepackage{mathrsfs}
\usepackage{amsmath,amssymb,amsthm}
\begin{document}
\title{No evidence for dynamical dark energy in two models}

\author{Deng Wang}

\email{Cstar@mail.nankai.edu.cn}
\affiliation{Theoretical Physics Division, Chern Institute of Mathematics, Nankai University,
Tianjin 300071, China}
\author{Xin-He Meng}
\affiliation{{Department of Physics, Nankai University, Tianjin 300071, China}}
\begin{abstract}
To investigate whether the dark energy evolves over time, we propose two null tests and constrain them using the data combination of cosmic microwave background radiation, baryonic acoustic oscillations, Type Ia supernovae, Planck-2015 lensing and cosmic chronometers. We find that, for these two models, there is no evidence of the dynamical dark energy at the $1.2\sigma$ confidence level. Interestingly, both models could slightly alleviate (i) the current Hubble constant ($H_0$) tension between the global fitting derivation by the Planck collaboration and the local observation by Riess {\it et al.}; (ii) the root-mean-square density fluctuations ($\sigma_8$) tension between the Planck-2015 data and several low-redshift large scale structure probes.

\end{abstract}
\maketitle
\section{Introduction}
During the past two decades, a large number of cosmological observations such as Type Ia supernovae (SNIa) \cite{1,2},  baryonic acoustic oscillations (BAO) \cite{3}, cosmic microwave background (CMB) radiation \cite{4,5} and weak gravitational lensing \cite{6} have confirmed that out universe is undergoing an accelerated expansion phase. The discovery of cosmic acceleration, in the framework of general relativity plus basic cosmological principle and perfect fluid assumption, has established the standard cosmological paradigm, namely the cosmological constant and cold dark matter ($\Lambda$CDM) model. This model has been verified to be very successful in describing various phenomena, from the origin and evolution of large scale structure (LSS) to the late-time acceleration. Especially, the Planck-2015 public release with an unprecedented accuracy has demonstrated, once again, the correctness of the standard six-parameter $\Lambda$CDM cosmology \cite{5}. Most recently, the galaxy clustering and weak gravitational lensing data from the first year (Y1) release of the Dark Energy Survey (DES) also proved its validity in characterizing the evolution of the universe \cite{7}. However, the $\Lambda$CDM model is not impeccable and faces several intractable problems: (i) the small scale crisis of CDM \cite{8}; (ii) the well-known coincidence and fine-tuning problems \cite{9}; (iii) the Hubble constant tension over between indirect global measurement by the Planck Collaboration \cite{10} under the assumption of $\Lambda$CDM and the direct local observation by Riess {\it et al.} \cite{11} using improved SNIa calibration techniques; (iv) the inconsistencies of the amplitude of matter density fluctuation between the Planck-2015 data and several low-redshift LSS probes including lensing, cluster counts and redshift space distortions (RSD) \cite{10,12,13}; (v) three new unknown entities are required: one which drove the inflation of the very early universe, another which acts as DM and the third which serves as DE. Meanwhile, it appears that there is no compelling reason to use only six parameters to describe the universe in light of very abundant observations \cite{14}. As a consequence, facing the above challenges, cosmologists have to question the validity of the $\Lambda$CDM paradigm. In general, they mainly propose two effective approaches to resolve the present problems and tensions: (i) in the framework of GR, one can implement a simple extension or modification to the $\Lambda$CDM model \cite{15,16,17,18,19,20,21,22,23,24,25,26,27,28,29,30,31}; (ii) while the GR breaks down at galactic or larger scales, one needs to modify the standard lagrangian of the Einstein's gravity \cite{32,33,34,35,36,37,38,39}.

To the best of our knowledge, the DE governing the background expansion of the universe is phenomenologically a cosmic fluid with an equation of state (EoS) $\omega_{de}\approx-1$, which violates the strong energy condition. Moreover, the DE fluid is homogeneously permeated in the universe and it has no the property of clustering unlike the DM. Except for the above characteristics, we are still unclear about the nature of DE such as its origin and constituent. It is worth noting that the important question whether the DE is dynamical is always argued by many authors in recent years \cite{40,41,42,43,44,45,46,47,48,49,50,51,52,53,54,55,56,57,58}. Recently, in light of the recent observations, Zhao \textit{et al.} \cite{59} made a new progress on studying the nature of DE. They claimed that the dynamical dark energy (DDE) is preferred over the constant DE ($\Lambda$CDM) from the point of view of cosmological fit alone at the 3.5$\sigma$ confidence level (CL), although the Bayesian evidence for the DDE is insufficient to favor it over $\Lambda$CDM. This implies that this important question is still in suspense and needs to be further investigated. Considering that the Sloan Digital Sky Survey (SDSS) IV extended  Baryon Oscillation Spectroscopic Survey (eBOSS) data release 14 (DR14) quasar sample is publicly released \cite{60}, we propose two new null tests to study whether the DE is actually dynamical at all. We find that there is no evidence of the DDE at the $1.2\sigma$ CL for these two models.

The rest of this paper is outlined in the following manner. In the next section, we introduce two new null tests for the cosmological constant scenario.    In Section III, we describe the observational data sets and analysis methodology, while we present our results in Section IV. The discussions and conclusions are presented in the final section.

\section{Null tests}
The Friedmann equations, the conservation equation of stress-energy tensor, and the EoS compose a close dynamical system to characterize the background evolution of the universe. For a Friedmann-Robertson-Walker (FRW) universe, the time-component Friedmann equation and the energy conservation one can be, respectively, written as
\begin{equation}
\frac{\dot{a}^2}{a^2}=\frac{\rho}{3}, \label{1}
\end{equation}
\begin{equation}
\dot{\rho}+3\frac{\dot{a}}{a}(\rho+p)=0, \label{2}
\end{equation}
where $a$, $p$ and $\rho$ denote the scale factor, pressure and energy density of the cosmic fluid, respectively, and the dot is the derivative with respect to the cosmic time $t$. It is noteworthy that we take the units $8\pi G=c=\hbar=1$ throughout this paper. To study that the DE is dynamical or not, we propose the following two DE density parameterizations
\begin{equation}
\rho_{de1}=\rho_{de0}(1+z)^{\alpha}=\rho_{de0}a^{-\alpha}, \label{3}
\end{equation}
\begin{equation}
\rho_{de2}=\rho_{de0}(1+\beta\frac{z}{1+z})=\rho_{de0}[1+\beta(1-a)], \label{4}
\end{equation}
where $z$, $\rho_{de0}$, $\alpha$ and $\beta$ denote the redshift, present DE density and free parameters of two DE models. One can easily find that these two models reduce to the $\Lambda$CDM case $\rho_{de}=\rho_{de0}$, when $\alpha=\beta=0$. The exact values of $\alpha$ and $\beta$ will be obtained by confronting both models with the latest cosmological observations. Inserting Eqs. (\ref{3}-\ref{4}) into Eq. (\ref{2}), the DE pressures of two null test scenarios are expressed as
\begin{equation}
p_{de1}=(-1+\frac{\alpha}{3})\rho_{de0}a^{-\alpha}, \label{5}
\end{equation}
\begin{equation}
p_{de2}=\rho_{de0}[\frac{4}{3}\beta a-(1+\beta)], \label{6}
\end{equation}
where $p_{de}$ denotes the DE pressure. One can find that the terms $\frac{\alpha}{3}\rho_{de0}a^{-\alpha}$ in Eq. (\ref{5}) and $\beta\rho_{de0}(\frac{4}{3} a-1)$ in Eq. (\ref{6}) represent the corrections of our null tests to the $\Lambda$CDM case, where $p_{de}=-\rho_{de0}$. Combining Eq. (\ref{1}) with Eqs. (\ref{3}-\ref{4}), the dimensionless Hubble parameters $E(a)$ of both models can be shown as
\begin{equation}
E_1(a)=\left[\Omega_ma^{-3}+(1-\Omega_m)a^{-\alpha}\right]^{\frac{1}{2}},   \label{7}
\end{equation}
\begin{equation}
E_2(a)=\left[\Omega_ma^{-3}+(1-\Omega_m)(1+\beta-\beta a)\right]^{\frac{1}{2}},   \label{8}
\end{equation}
where $\Omega_m$ is the dimensionless matter density parameter. Since we focus mainly on the late-time cosmology, we neglect the contribution from the radiation ingredient in the cosmic pie. Considering that the Planck CMB data has given a very tight constraint on the present cosmic curvature $\Omega_{k}<|0.005|$ \cite{61}, we also neglect the contribution from curvature to the evolution of the universe. Subsequently, using Eqs. (\ref{3}-\ref{4}) and Eqs. (\ref{5}-\ref{6}), we obtain the effective EoS of DE $\omega_{de}(a)$ of two models as
\begin{equation}
\omega_{de1}(a)=-1+\frac{\alpha}{3},   \label{9}
\end{equation}
\begin{equation}
\omega_{de2}(a)=-1+\frac{\beta a}{3(1+\beta-\beta a)}.   \label{10}
\end{equation}
In order to perform constraints on these two null tests later on, it is necessary to discuss the possible ranges of $\alpha$ and $\beta$. Assuming $-3<\omega_{de}<1$, we obtain $-6<\alpha<6$ for the first model (M1) and $-2<\frac{\beta a}{3(1+\beta-\beta a)}<2$ for the second one (M2). Subsequently, because we are interested in the evolution of the late universe, we also derive $-6<\beta<6$ by taking $a=1$.

In addition, we consider the linear perturbations of background metric. The general scalar mode perturbation of FRW background spacetime can be shown as \cite{62,63,64}
\begin{equation}
ds^2=-(1+2\Phi)dt^2+2a\partial_iBdtdx+a^2[(1-2\Psi)\delta_{ij}+2\partial_i\partial_jE]dx^idx^j,  \label{11}
\end{equation}
where $\Phi$ and $\Psi$ denote the linear gravitational potentials. Following Ref. \cite{63} and using the synchronous gauge $\Phi=B=0$, $\Psi=\eta$ and $E=-(h+6\eta)/2k^2$, the energy-momentum conservation equations for the cosmic fluid in the the synchronous gauge can be expressed as
\begin{equation}
\delta'=-(1+\omega)(\theta+\frac{h'}{2})-3(\frac{\delta p}{\delta\rho}-\omega)\mathcal{H}\delta, \label{12}
\end{equation}
\begin{equation}
\theta'=(3\omega-1)\mathcal{H}\theta-\frac{\omega'}{1+\omega'}\theta+\frac{\delta p}{\delta\rho}\frac{k^2\delta}{1+\omega}-k^2\delta, \label{13}
\end{equation}
where $\sigma$, $\delta$, $\theta$ and $\mathcal{H}$ denote, respectively, the shear, density perturbation, velocity perturbation and conformal Hubble parameter, and the prime is the derivative with respect to the conformal time. Furthermore, the DE perturbations are shown as
\begin{equation}
\delta_{de}'=-(1+\omega_{de})(\theta_{de}+\frac{h'}{2})-3\mathcal{H}\omega'_{de}\frac{\theta_{de}}{k^2}+3\mathcal{H}(\omega_{de}-c_s^2)[\delta_{de}+3\mathcal{H}(1+\omega_{de})\frac{\theta_{de}}{k^2}],   \label{14}
\end{equation}
\begin{equation}
\theta_{de}'=(3c_s^2-1)\mathcal{H}\theta_{de}+\frac{c_s^2}{1+\omega_{de}}k^2\delta_{de}, \label{15}
\end{equation}
where $c_s^2$ denotes the physical sound speed (SS) in the rest frame. For the purpose to avoid the unphysical SS, we have adopted $c_s^2=1$ in the following numerical analysis. Meanwhile, for the convenience of calculations, we also take $\sigma=0$. Note that since there is no interaction between DM and DE in the dark sector in our analysis, the perturbations of these two components follow independently the standard evolution formula presented in \cite{62,63,64}.

\begin{figure}
\centering
\includegraphics[scale=0.45]{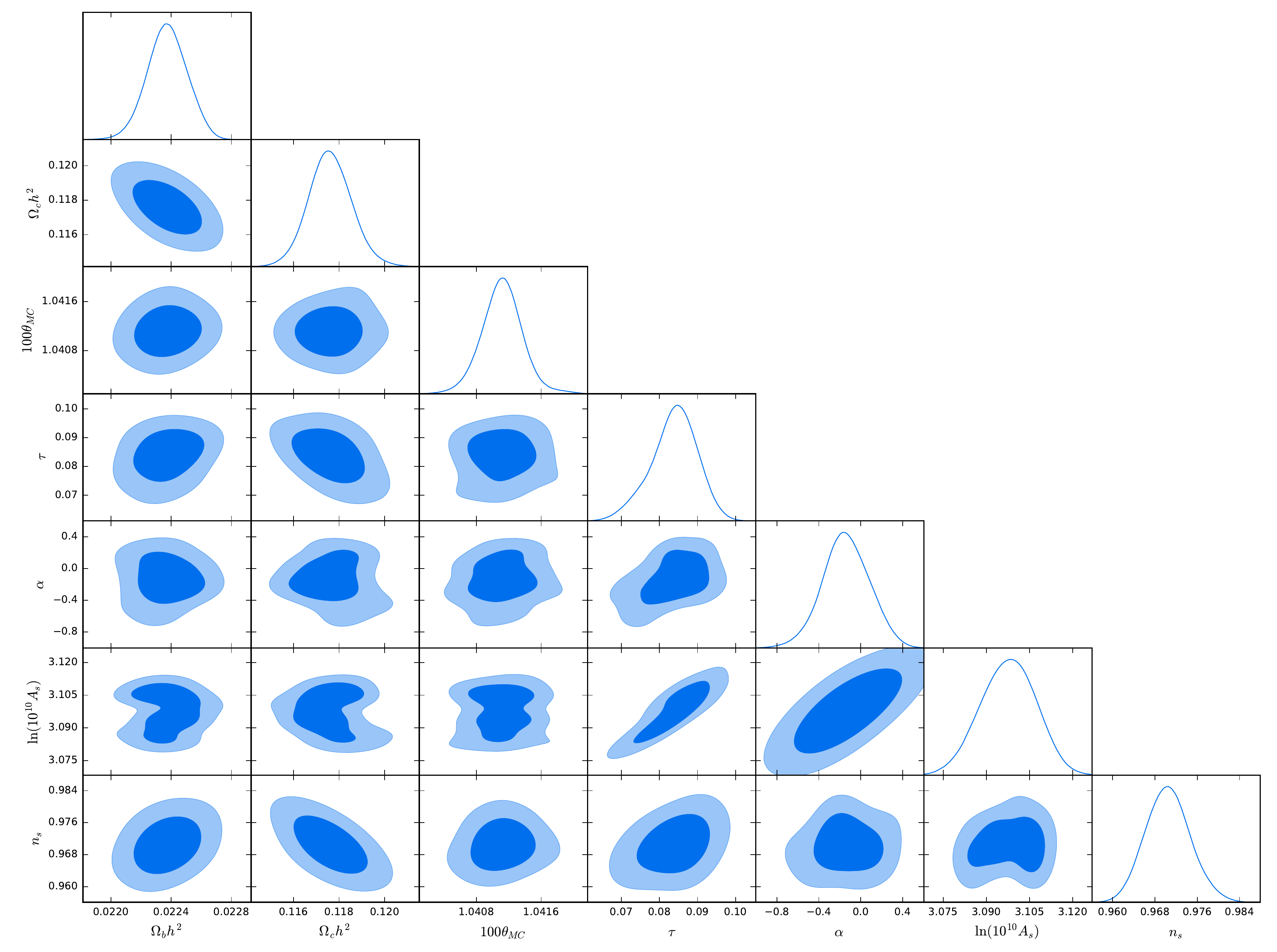}
\caption{The $68\%$ and $95\%$ confidence regions of the M1 using the data combination CBSLC.}\label{f1}
\end{figure}

\begin{figure}
\centering
\includegraphics[scale=0.45]{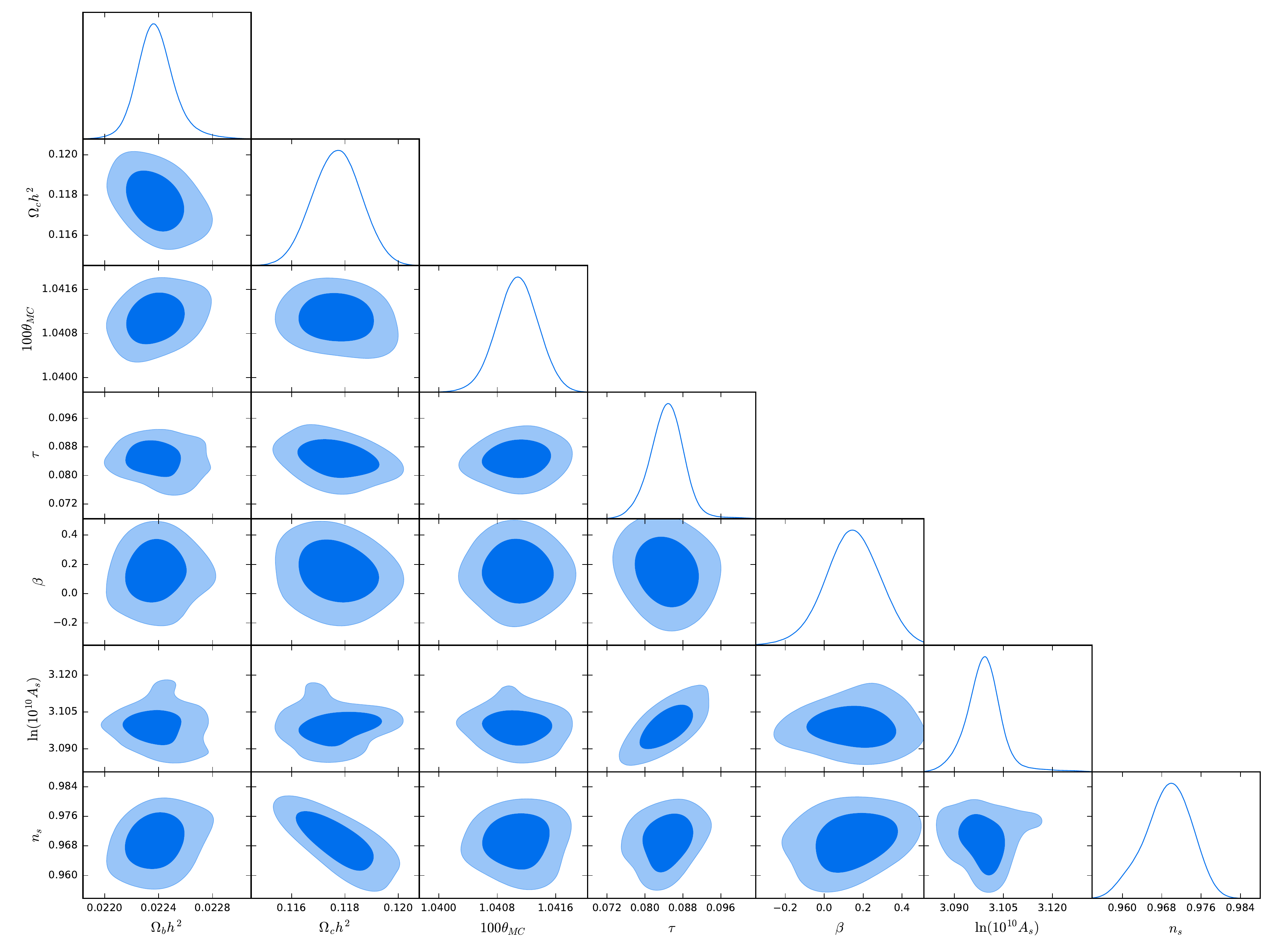}
\caption{The $68\%$ and $95\%$ confidence regions of the M2 using the data combination CBSLC.}\label{f2}
\end{figure}

\begin{figure}
\centering
\includegraphics[scale=0.6]{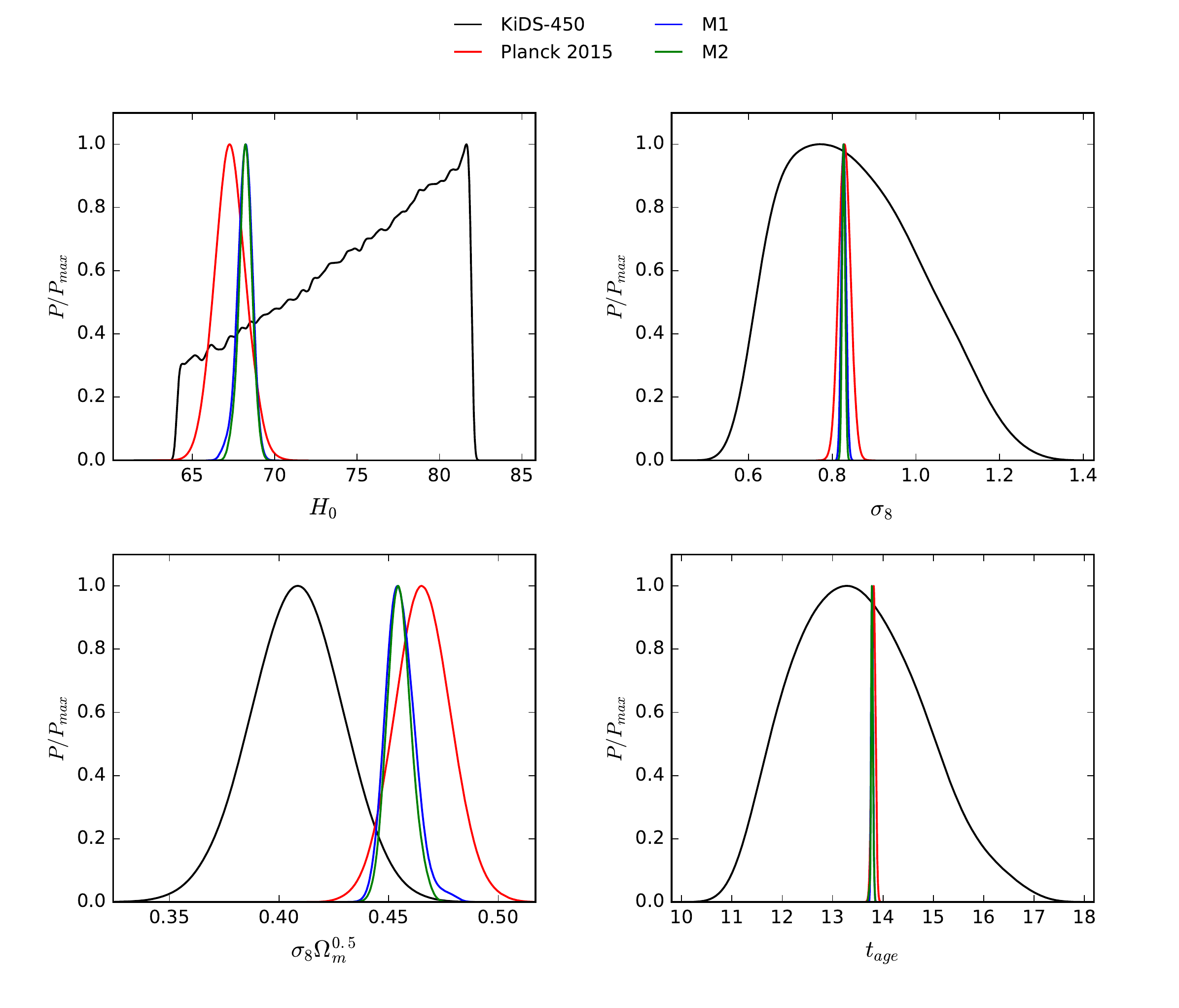}
\caption{The comparison between the 1-dimensional marginalized posterior distributions of $H_0$, $\sigma_8$, $\sigma_8\Omega_m^{0.5}$ and $t_{age}$ from the M1 and M2 using the data combination CBSLC and those from KiDS-450 and Planck-2015 results under the assumption of $\Lambda$CDM. }\label{f3}
\end{figure}

\begin{figure}
\centering
\includegraphics[scale=0.55]{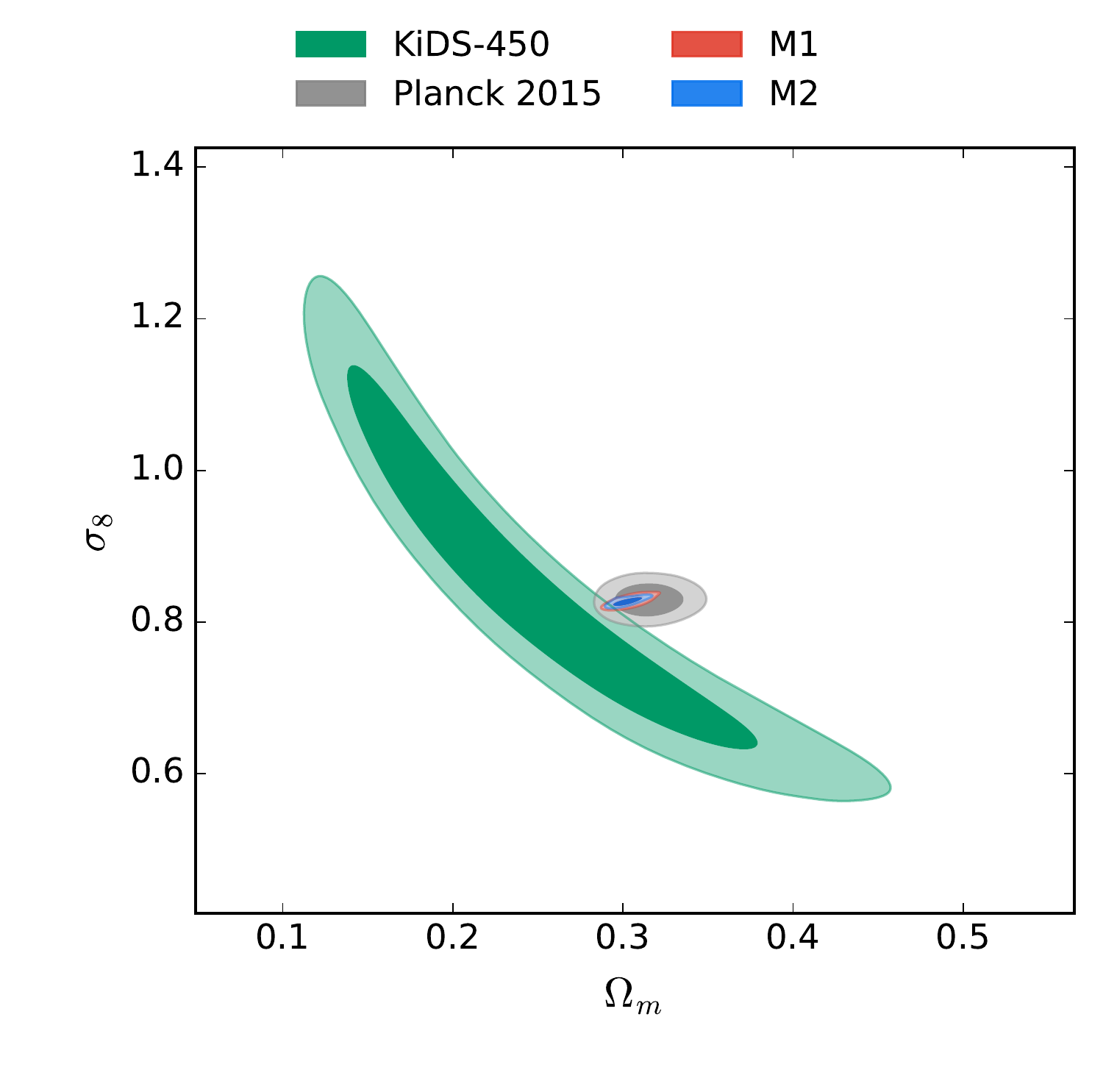}
\includegraphics[scale=0.55]{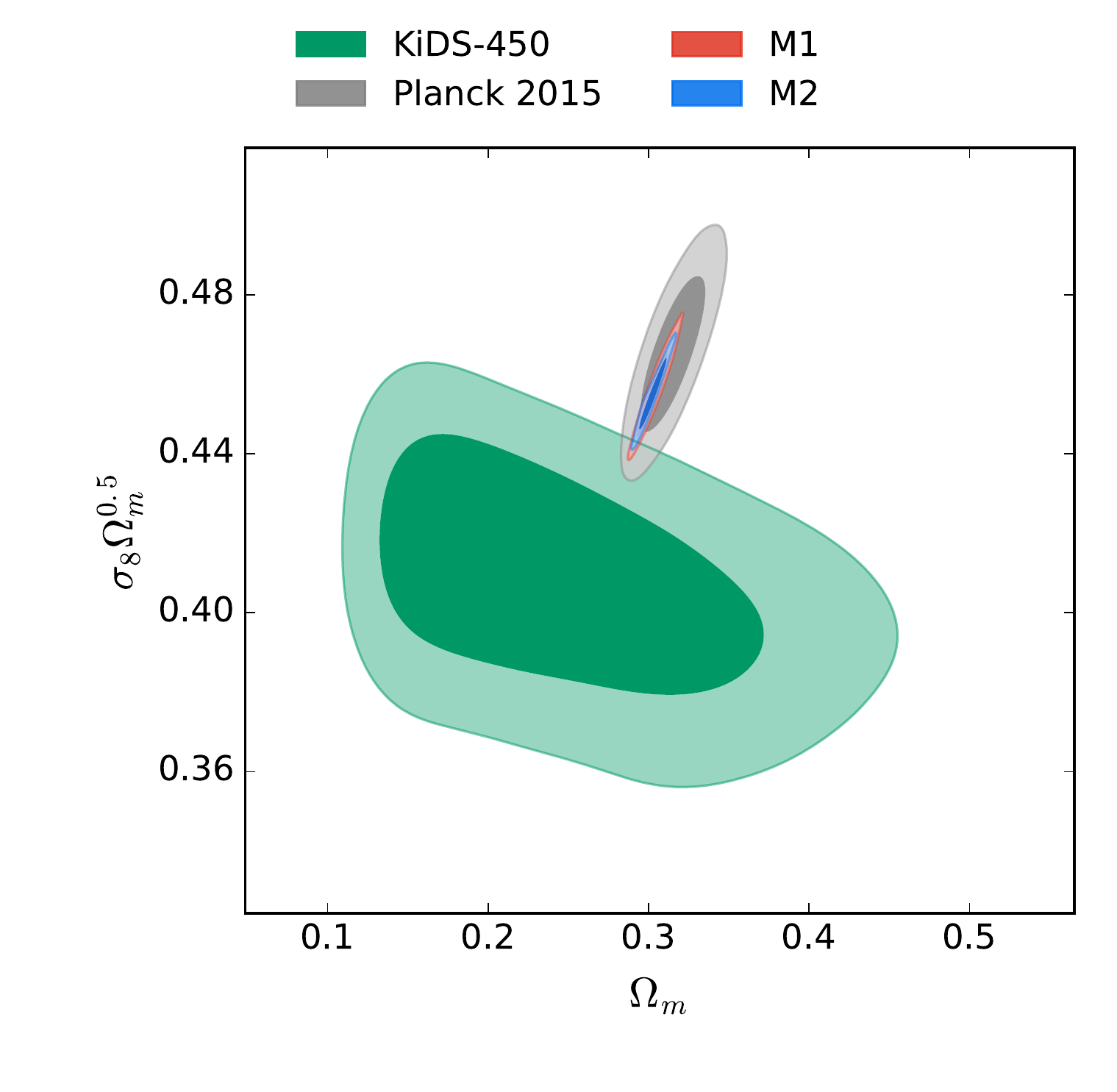}
\caption{The comparison between the 2-dimensional confidence regions of the M1 and M2 using the data combination CBSLC and those of KiDS-450 and Planck-2015 results under the assumption of $\Lambda$CDM in the planes of $\Omega_m-\sigma_8$ and $\Omega_m-\sigma_8\Omega_m^{0.5}$.}\label{f4}
\end{figure}

\begin{figure}
\centering
\includegraphics[scale=0.5]{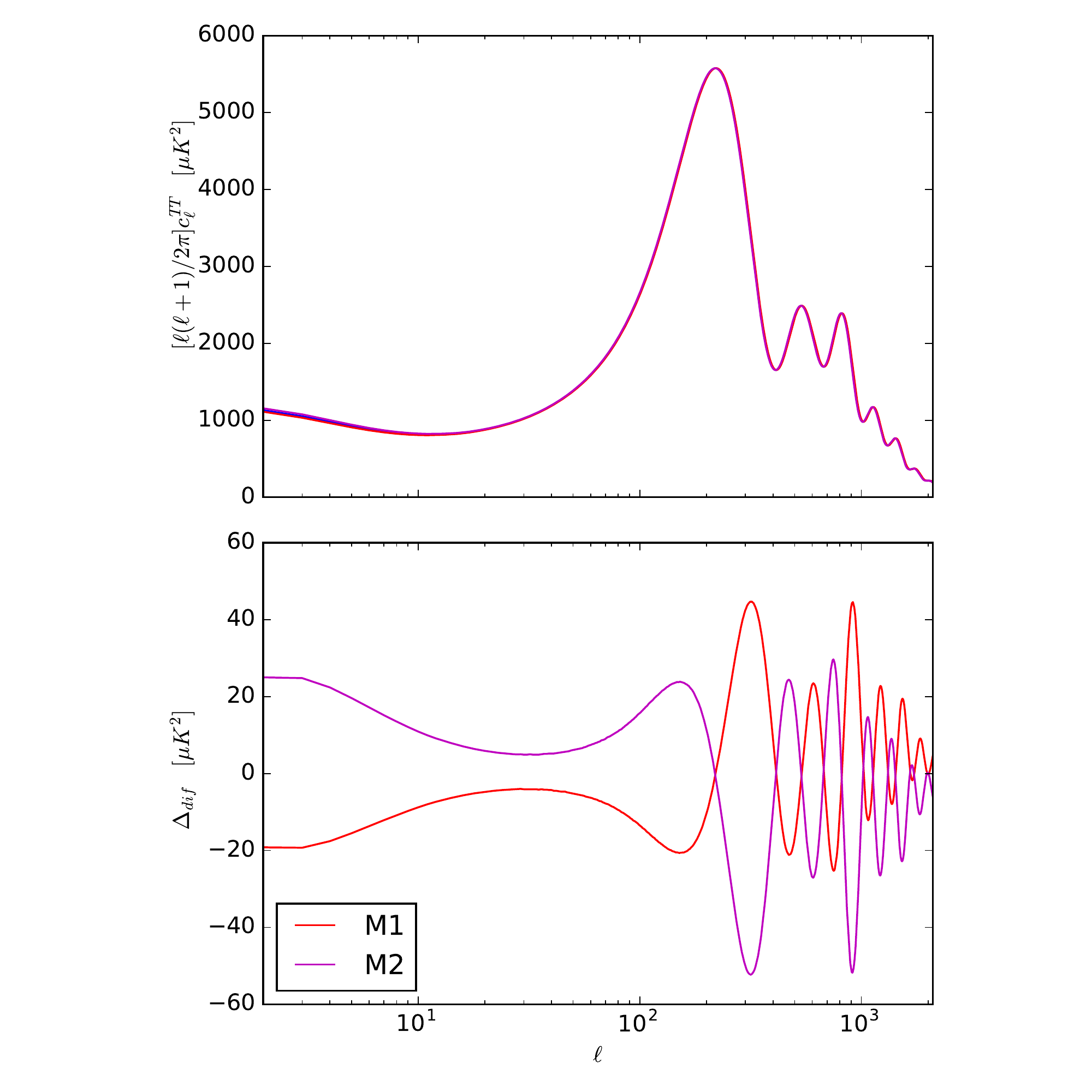}
\caption{The CMB temperature power spectrums of the $\Lambda$CDM model (blue), M1(red) and M2 (purple) as well as the relative differences $\Delta_{dif}$ between the $\Lambda$CDM and two models. }\label{f5}
\end{figure}

\begin{figure}
\centering
\includegraphics[scale=0.5]{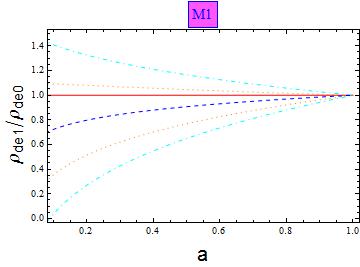}
\includegraphics[scale=0.5]{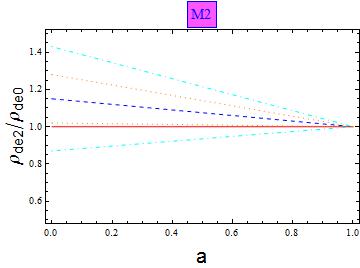}
\includegraphics[scale=0.5]{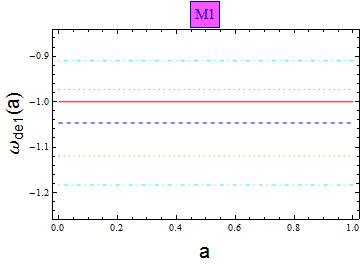}
\includegraphics[scale=0.5]{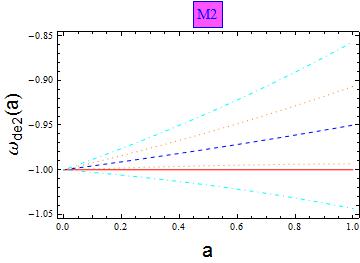}
\includegraphics[scale=0.5]{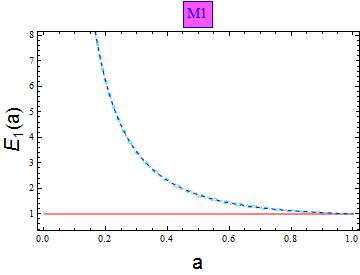}
\includegraphics[scale=0.5]{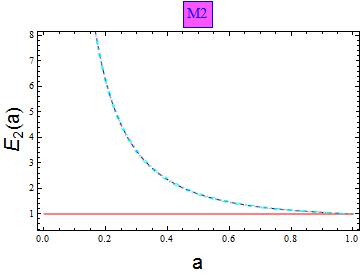}
\caption{The relations between the scale factor $a$ and the relative DE density $\rho_{de}/\rho_{de0}$, EoS of DE $\omega_{de}(a)$ and dimensionless Hubble parameter $E(a)$, respectively. The red lines correspond to the $\Lambda$CDM model $\rho_{de}/\rho_{de0}=1$, $\omega_{de}=-1$ and a reference line $E(a)=1$ in the upper, medium and lower panels, respectively. The blue (dashed), orange (dotted) and cyan (dash-dotted) lines correspond to the mean values of $\alpha$ (M1) or $\beta$ (M2), $1\sigma$ and $2\sigma$ confidence regions, respectively.}\label{f6}
\end{figure}

\section{Data and methodology}
In this section, to investigate whether the DE is time-dependent at all, we use the latest cosmological observations to constrain the above two null test models. Their corresponding parameter spaces can be expressed as
\begin{equation}
\mathbf{P_1}=\{\Omega_bh^2, \quad \Omega_ch^2, \quad 100\theta_{MC}, \quad \tau, \quad \alpha, \quad  \mathrm{ln}(10^{10}A_s), \quad  n_s \},   \label{16}
\end{equation}
\begin{equation}
\mathbf{P_2}=\{\Omega_bh^2, \quad \Omega_ch^2, \quad 100\theta_{MC}, \quad \tau, \quad \beta, \quad  \mathrm{ln}(10^{10}A_s), \quad  n_s \},   \label{17}
\end{equation}
where $\Omega_bh^2$ and $\Omega_ch^2$ denote the present baryon and CDM densities, $\theta_{MC}$ is the ratio between the angular diameter distance and sound horizon at the redshift of last scattering $z_\star$, $\tau$ is the optical depth due to reionization, $\alpha$ and $\beta$ refer to free parameters of two null tests scenarios, $\mathrm{ln}(10^{10}A_s)$ and $n_s$ are the amplitude and spectral index of primordial power spectrum at the pivot scale $K_0=0.05$ Mpc$^{-1}$. It is noteworthy that $h$ is related to the Hubble constant $H_0$ by $h\equiv H_0/(100\, \mathrm{kms^{-1}Mpc^{-1}})$.

The observational data sets used in this paper can be exhibited as follows:

\textit{CMB}: With a high accuracy, the Planck-2015 CMB data has measured the topology, matter constituents, LSS formation, evolution of the universe. Here we employ the CMB temperature and polarization data from the full Planck survey \cite{61}, which includes the large angular-scale temperature and polarization anisotropy measured by the Planck LFI experiment and the small-scale anisotropies measured by the Planck HFI one. More specifically, this data set consists of the likelihoods of temperature at $30\leqslant \ell\leqslant 2500$, the cross-correlation of temperature and polarization, the polarization power spectra, and the low-$\ell$ temperature and polarization likelihood at $2\leqslant \ell\leqslant 29$.

\textit{BAO}: BAO observations are geometric and, to a large extent, unaffected by errors in the nonlinear evolution of the matter density field and other systematic errors which may affect other astrophysical measurements. Measuring the position of these oscillations in the matter power spectra at different redshifts can constrain the expansion history of the universe after decoupling, consequently breaking degeneracies in the interpretation of CMB anisotropies.
In this paper, we use four BAO measurements: the 6dFGS sample at effective redshift $z_{eff}=0.106$ \cite{65}, the SDSS-MGS one at $z_{eff}=0.15$ \cite{66}, and the LOWZ at $z_{eff}=0.32$ and CMASS at $z_{eff}=0.57$ data from the SDSS-III BOSS DR12 sample \cite{67}. As mentioned above, we also include the latest SDSS-IV eBOSS DR14 quasar sample in our analysis \cite{60}.

\textit{SNIa}: SNIa are substantially power probes of cosmology and are considered as standard candles in exploring the evolution of the universe, particularly, the EoS of DE. We employ the largest SNIa sample `` Joint Light-curve Analysis '' (JLA) to date constructed from SNLS and SDSS data, together with several low-redshift SNIa data \cite{68}.

\textit{Lensing}: As a complementary probe, we also include the Planck-2015 lensing likelihood in our analysis \cite{69}. The gravitational lensing by the LSS leaves imprints on the CMB temperature and polarization which could be observed in high angular resolution, low noise observations, such as those from the Planck survey.

\textit{Cosmic Chronometers}: The Cosmic Chronometers (CC) observations are determined by using the most massive and passively evolving galaxies based on the `` galaxy differential age ''. Note that this kind of observational Hubble parameter data is model-independent. In this analysis, we adopt 30 CC measurements covering the redshift range $z\in[0.07, 1.97]$ to constrain the above two DE models \cite{70}.

We employ the the Markov Chain Monte Carlo (MCMC) technique to infer the posterior probability distributions of different model parameters. We modifies carefully the online MCMC package CosmoMC \cite{71}, which obeys a convergence diagnostic based on the Gelman and Rubin statistic, and Boltzmann code CAMB \cite{72}. To implement the standard Bayesian analysis, we adopt the prior ranges for different parameters in the following manner: $\Omega_bh^2 \in [0.005, 0.1]$, $\Omega_ch^2 \in [0.001, 0.99]$, $100\theta_{MC} \in [0.5, 10]$, $\tau \in [0.01, 0.8]$, $\mathrm{ln}(10^{10}A_s) \in [2, 4]$, $n_s \in [0.8, 1.2]$, $\alpha \in [-3, 3]$, $\beta \in [-3, 3]$. Since the range $[-6, 6]$ is too wide for $\alpha$ and $\beta$, we choose the relatively small range $[-3, 3]$ for them. Subsequently, in order to carry out the strictest constraint on the cosmological parameters, we use a data combination of CMB + BAO + SNIa + Lensing + CC, which is labelled as `` CBSLC '' in the following analysis.

\begin{table}[h!]
\renewcommand\arraystretch{1.3}

\caption{The mean values with corresponding 68$\%$ limits and best fit of different model parameters in the M1 and M2 using the data combination CBSLC. }
\label{t1}
\begin{tabular} { l c c c c }

\hline
\hline
Parameters & \quad Mean with errors (M1) & \quad Best fit (M1) & \quad Mean with errors (M2) & \quad Best fit (M2)\\
\hline
{$\Omega_b h^2   $}        & $0.02237\pm 0.00013        $& 0.02238 & $0.02238^{+0.00012}_{-0.00014}$ & 0.02235 \\

{$\Omega_c h^2   $}      & $0.11761\pm 0.00093        $
 & 0.11731 & $0.11770\pm 0.00088        $ & 0.11672 \\

{$100\theta_{MC} $}     & $1.04112\pm 0.00025        $ & 1.04105 & $1.04108\pm 0.00028        $ & 1.0411 \\

{$\tau           $}  & $0.0839^{+0.0068}_{-0.0052}$ & 0.0869 & $0.0847\pm 0.0035          $ & 0.0845
                                                        \\

{${\rm{ln}}(10^{10} A_s)$}     & $3.0977\pm 0.0084          $ & 3.103  & $3.0992\pm 0.0057          $ & 3.0971   \\

{$n_s            $}         & $0.9704\pm 0.0041          $ & 0.9715 & $0.9693^{+0.0053}_{-0.0042}$ & 0.9722 \\

{$\alpha        $}  & $-0.14\pm 0.22             $ & -0.09  &  ---  &  ---                                                \\

{$\beta          $}  & --- &  --- & $0.15\pm 0.13              $ & 0.28                                                  \\

\hline
$H_0              $      & $68.18\pm 0.40             $ & 68.36 & $68.23\pm 0.41             $  & 68.56                                                   \\

$\Omega_m              $ & $0.3017\pm 0.0053          $ & 0.3003         & $0.3023\pm0.0051$  & 0.2972 \\

$\sigma_8              $ & $0.8266^{+0.0044}_{-0.0054}$ & 0.8280 & $0.8272\pm 0.0035          $ & 0.8239 \\

$\sigma_8\Omega_m^0.5              $ & $0.4554^{+0.0058}_{-0.0072}$ & 0.4537 & $0.4549\pm 0.0058          $ & 0.4492 \\

$t_{age}              $      & $13.784^{+0.018}_{-0.020}$ & 13.781 & $13.782\pm0.020$  & 13.779                                                   \\

\hline
\hline
\end{tabular}
\end{table}

\section{Results}
Taking advantage of the combined data sets CBSLC, our numerical results are presented in Tab. \ref{t1}, which includes the mean values with their 68$\%$ limits and best fit of different model parameters in the M1 and M2. The one-dimensional marginalized posterior distributions and two-dimensional contours for both models are presented in Figs. \ref{f1}-\ref{f2}. For the key parameters $\alpha$ (M1) and $\beta$ (M2), we also give the 95$\%$ limits $\alpha=-0.14\pm 0.22 (1\sigma)\pm 0.41 (2\sigma)$ and $\beta=0.15\pm 0.13 (1\sigma)\pm 0.28 (2\sigma)$. It is easy to find that the M1 is consistent with the prediction of $\Lambda$CDM at the $1\sigma$ CL. However, this is not the case for the M2, which prefers mildly a small positive value of $\beta$ at the $1\sigma$ CL implying the hints of DDE. Actually, the M2 is compatible with the $\Lambda$CDM model at the $1.2\sigma$ CL. Aa a consequence, for two null tests, we can conclude that there is no evidence of the DDE at the $1.2\sigma$ CL (this is the main conclusion of this paper). Subsequently, we find that the values of spectral index $n_s$ of both models are in a good agreement with the Planck-2015 estimation $n_s=0.9655\pm0.0062$ at the $1\sigma$ CL (Planck TT + lowP) \cite{61}, and that the scale-invariant Harrison-Zeldovich-Peebles (HZP) power spectrum ($n_s=1$) \cite{73,74,75} is still strongly excluded at the 7.22$\sigma$ and 5.79$\sigma$ CL in the M1 and M2, respectively. Meanwhile, we obtain the minimal value of $\chi^2$ of M1 and M2 as 13715.638 and 13713.643, respectively, and find that the M2 gives a better cosmological fit than the M1 by a difference $\Delta\chi^2=1.995$.

Considering the recent cosmic shear analysis of $\sim$450 deg$^2$ of imaging data from the Kilo Degree Survey (KiDS-450) \cite{76} and Planck CMB data analysis \cite{61}, we make a comparison between their results and the predictions of our two models (see Figs. \ref{f3}-\ref{f4}). Combining Tab. \ref{t1} with Fig. \ref{f3}, we find that the M1 and M2 could slightly alleviate, respectively, the current $H_0$ tension from 3.4$\sigma$ to 2.83$\sigma$ and 2.80$\sigma$ between the global measurement by Planck and the local observation by Riess \textit{et al.} using the data combination CBSLC. As noted in \cite{76}, the KiDS-450 analysis is not particularly sensitive to $H_0$ so that constraint on it are relatively loose. Interestingly, one can also find that the similar consequence occurs in the constraint on the age of the universe $t_{age}$ by KiDS-450, which gives $t_{age}=13.5^{+1.1}_{-1.5}$ Gyr. The constraints on $t_{age}$ provided by both models is well consistent with the prediction $t_{age}=13.813\pm0.038$ Gyr of Planck at the 1$\sigma$ CL (Planck TT + lowP) \cite{61}. Subsequently, we find that the values of the amplitude of matter density fluctuation $\sigma_8$ from two models are compatible with those of KiDS-450 and Planck at the 1$\sigma$ CL (see the upper right panel of Fig. \ref{f3} and left one of Fig. \ref{f4}). Due to the fact that the KiDS-450 collaboration finds a 2.3$\sigma$ tension for the composite parameter $S_8=\sigma_8\sqrt{\Omega_m/0.3}$ between KiDS-450 and Planck-2015 results \cite{76}, we calculate the combination $\sigma_8\Omega_m^{0.5}$ and also obtain a 2.3$\sigma$ tension between these two surveys. Since the constraint on $\sigma_8$ from KiDS-450 is much looser than the left three cases and we cannot identify the LSS information very well, we exhibit the relation between the combination $\sigma_8\Omega_m^{0.5}$ and $\Omega_m$ in the right panel of Fig. \ref{f4}. We find that this tension can be mildly alleviated from 2.3$\sigma$ to 2.14$\sigma$ and $2.15\sigma$ in the M1 and M2, respectively. Furthermore, we also investigate the effects of modified dark sector from our two models on the CMB temperature power spectrum, and find that the M1 and M2 give almost the same prediction as $\Lambda$CDM at small and large scales and that the relative differences $\Delta_{dif}$ between the $\Lambda$CDM and two models are too small to affect hardly the properties of dark sector of the universe (see Fig. \ref{f5}).

In addition, we are of much interest in studying the late-time background evolution of our two null tests. In the two upper panels of Fig. \ref{f6}, we find that the relative DE densities of the M1 and M2 tend to infinitely approach the standard cosmological model and their confidence regions tend to converge into the $\Lambda$CDM model at the present epoch. The EoS of DE of M1 remains a constant $\omega_{de1}=-1.046\pm0.073$, which is very compatible with the Planck-2015 analysis $\omega_{de}=-1.006\pm0.045$ at the $1\sigma$ CL \cite{61}, and that of M2 tends to deviate slowly from the $\Lambda$CDM model at low redshifts but still keeps consistent with the prediction of $\Lambda$CDM at the 1.2$\sigma$ CL (see the two medium panels of Fig. \ref{f6}). In the two lower panels, from the view of expansion rate of the universe, we find that both models cannot be distinguished from the $\Lambda$CDM one and share the same evolutional behavior of the universe at late times.

\section{Discussions and conclusions}
There is no doubt that our universe is experiencing an accelerated expansion today. However, we still know little about the nature of the accelerated mechanism (or DE physics). With more and more high-precision astronomical data, understanding better the underlying physics of DE phenomena is not only an urgent task but also a large challenge to the modern cosmology. One important question about the DE issue is whether the DE actually evolves with time at all. Considering the recent public release of the SDSS-IV eBOSS DR14 quasar sample, we propose two null tests in combination with other data sets including CMB, BAO, SNIa and CC, to investigate that the DE is dynamical or not.

Using the tightest constraint CBSLC we can provide, we obtain the following conclusions: (i) For both models, there is no evidence of the DDE at the $1.2\sigma$ CL; (ii) The scale invariance of HZP primordial power spectrum is strongly excluded, while their constrained values of spectral index are in good agreement with the Planck analysis \cite{61}; (iii) The M1 and M2 could slightly alleviate, respectively, the current $H_0$ tension from 3.4$\sigma$ to 2.83$\sigma$ and 2.80$\sigma$ between the global measurement by Planck and the local observation by Riess \textit{et al.}; (iv) Making use of the composite LSS parameter $\sigma_8\Omega_m{0.5}$, we find that the $\sigma_8$ tension could also be moderately alleviated from 2.3$\sigma$ to 2.14$\sigma$ and $2.15\sigma$ in the M1 and M2, respectively; (v) Through analyzing the CMB temperature power spectrum and evolutional behaviors of cosmological quantities of these two models, we find that they just deviate very slightly from the $\Lambda$CDM model at the late universe, and that although these extremely small deviations affect hardly the properties of the dark sector of our universe, they may help us to resolve the current cosmological puzzles.

It is interesting that the EoS of DE of M2 has the same analytical expression as that of M1 when $a=1$, i.e., $\omega_{de2}(1)=-1+\beta/3$ (see also Eqs. (\ref{9}-\ref{10})). Nonetheless, since the free parameters $\alpha$ and $\beta$ from the constraint CBSLC have different exact values and uncertainties, there exists a very small difference between the current EoS of DE of both models.

Note that our conclusions and previous works by other authors are all limited to the understanding of systematics of each set of cosmological data. Meanwhile, to understand the systematics of different data sets is also an intriguing topic to be explored in the future.

It is also worth noting that we do not study the possibility whether using our two new null tests could also relieve the internal tensions of Planck CMB data, i.e., the so-called $\tau$ and $A_{lens}$ tensions \cite{10}, where $A_{lens}$ denotes the amplitude of lensing power relative to the physical value. This issue will be further discussed in the forthcoming work.

With gradually mounting astronomical data, we expect that future high-precision cosmological experiments can shed light on the properties of dark sector of the universe. Moreover, with the coming era of gravitational-wave astronomy, we also expect that the combination of two informational channels, gravitational sirens and electromagnetic signals, can help us explore the nature of DE better.

\section*{Acknowledgements}
D. Wang thanks Y. Liu, L. Xu, I. Brevik, W. Yang, Y. Li and Y. S for helpful discussions on cosmology and gravitational theories. X. Meng thanks B. Ratra and S. D. Odintsov for useful communications on data analysis and classical gravity.

\end{document}